\documentclass[useAMS,usenatbib,sort,final]{mn2e}
\usepackage{xspace}
\usepackage[utf8]{inputenc}

\usepackage[nointegrals]{wasysym}

\renewcommand\sun{\hbox{$\odot$}}
\renewcommand\earth{\hbox{$\oplus$}}

\usepackage{amsmath}

\RequirePackage{ifpdf}
   \ifpdf
      \usepackage[pdftex]{graphicx}
      \else
      \usepackage{graphicx}
     \fi

\newcommand{\corot}{COROT}
\newcommand{\mc}{\ensuremath{M_\mathrm{c}}\xspace}
\newcommand{\mtot}{\ensuremath{M_\mathrm{tot}}\xspace}
\newcommand{\pc}{\ensuremath{P_\mathrm{c}}\xspace}
\newcommand{\mdot}{\ensuremath{\dot M}\xspace}

\newcommand{\msun}{\ensuremath{\mathrm{M_{\astrosun}}}\xspace}

\newcommand{\lsun}{\ensuremath{\mathrm{L_{\astrosun}}}\xspace}
\newcommand{\mjup}{\ensuremath{\mathrm{M_{\jupiter}}}\xspace}
\newcommand{\mearth}{\ensuremath{\mathrm{M}_{\earth}}\xspace}

\title[Theoretical planetary mass spectra]{Theoretical planetary mass
  spectra -- a prediction for \corot}
\author[C. Broeg]{C. Broeg$^{1,2,3}$\thanks{E-mail:
broeg@space.unibe.ch}\\
$^{1}$ Physikalisches Institut, University of Bern, Sidlerstrasse 5, CH-3012 Bern, Switzerland \\
$^{2}$ Th\"uringer Landessternwarte,  Sternwarte 5,
Tautenburg, 07778, Germany\\
$^{3}$ Max-Planck Institute for extraterrestrial Physics,
Giessenbachstrasse, Garching, 85741, Germany}

\begin{document}
\date{Accepted xxx. Received xxx; in original form 2006 October}

\pagerange{\pageref{firstpage}--\pageref{lastpage}} \pubyear{2006}

\maketitle

\label{firstpage}

\begin{abstract}
The satellite \corot\ will search for close-in exo-planets around a few
thousand stars using the transit search method. The \corot\ mission holds the
promise of detecting numerous exo-planets. Together with radial
velocity follow-up observations,  the masses of the detected planets
will be known.\\
We have devised a method for predicting the expected planetary
populations and compared it to the already known exo-planets. Our
method works by looking at all hydrostatic envelope solutions of giant
gas planets that
could possibly exist in arbitrary planetary nebulae and comparing the
relative abundance of different masses. We have completed the first
such survey of hydrostatic equilibria in an orbital range covering
periods of 1
to 50 days.\\
Statistical analysis of the calculated envelopes suggests division into
three classes of giant planets that are distinguished by orbital
separation. We term them classes G (close-in), H, and J (large separation). Each class has distinct
properties such as a typical mass range. \\Furthermore, the division
between class H and J appears to mark important changes in the
formation: For close-in planets (classes G and H) the concept of a critical
core-mass  
is meaningless while it is important for class J. This result needs
confirmation by future dynamical analysis.
\end{abstract}


\section{Introduction}
\label{sec:introduction}
Since the discovery of 51-Peg\,b \citep{1995Natur.378..355M}, more
than 200 exo-planets have been discovered, most by the radial velocity
technique. This year, the satellite mission \corot\ \citep{corot}  will
be launched 
hoping to add many more planets to the list. The \corot\ satellite will
be on a two-fold mission: It will do A)  astroseismology
\citep{1998IAUS..185..301B} and B) look for planets using the transit
search method
\citep{1984Icar...58..121B,2000ApJ...529L..45C,2000dais.conf..221R}. The
transit search programme hopes to find a relatively large number of
planets. It is the task of theoreticians to make a prediction beforehand.

The standard giant planet formation model is the so-called
core-accretion model as in \citet{mizuno1980}. In this
model, planet formation starts with sedimentation and coagulation of
the condensible material into small solid cores
\citep{1989Icar...77..330W,1993ARA&A..31..129L,2004ARA&A..42..549G}. As
this core grows, it becomes massive enough to gravitationally bind
some gas. Consequently, it acquires an
envelope of gas and dust. The evolution of this envelope has been
studied by many authors
\citep[e. g.][]{1996Icar..124...62P,1986Icar...67..391B,1990A&A...238...83W,1991Icar...91...39W,1991Icar...91...53W}.

It should be mentioned, that planets could also form as described
by the
gravitational instability scenario \citep{2002ApJ...576..462B}. Nevertheless,
today's planets are in better agreement with the core-accretion
scenario \citep[see][]{2005Sci...310..251S}. In this paper we work on
the basis of the core-accretion scenario.

A natural procedure when trying to predict the distribution of giant
planets is the statistical approach: Calculate the evolution of a
large number of randomly placed planetesimal "seeds", starting
with small planetesimals and letting them evolve to the final
planet. For each seed, the full evolution is calculated including core
growth, accumulation of envelope, migration, etc. \citep[see
e.g.][]{2006tafp.conf...24B}\citep{2005A&A...434..343A,2005ApJ...626L..57A,2005ApJ...622L.145A}. This
has the advantage, that a large number of processes can be included
into the algorithm. Furthermore, the physical evolution is modeled in
a natural way. However, there are disadvantages as well. First of all,
these calculations are computationally intensive and a very large
number of such calculations need to be done in order to gain a
statistically significant result. In addition, one needs to know the
exact environmental conditions in which the evolution of the seeds
should be calculated. This is a problem. Even in our own
solar system, the conditions during the time of formation are only
vaguely known. Nebula densities range between two extremes: There must
have been enough material to form all the planets  \citep[the concept of
the minimum mass solar nebula:
][]{1981PThPS..70...35H,1979E&PSL..43...22H,1977PASJ...29..163H} and
the nebula must be gravitationally stable. For a more thorough
discussion on the nebula variety see \citet{2000prpl.conf.1081W}. For other stars, the
primordial proto-planetary nebula is  constrained even less. As long as
these important parameters are not known with some degree of precision, we
think that it will be difficult to make a good prediction in this way.

Therefore, we use a different approach. We study  all possible
equilibrium  states consisting of a solid core and a gaseous
envelope with cores of different sizes and a range of nebula
densities.  

It is our goal to make a prediction for \corot. As this satellite
mission will only be sensitive to planetary orbits shorter than 50
days \citep{2003A&A...405.1137B}, we'll restrict our prediction to close-in planets ranging from 1
to 50 days orbital period. Our prediction is only valid for gas giant
planets, terrestrial planets' mass distributions cannot be predicted
in this way.

\section{Planet prediction method}
\label{sec:plan-pred-meth}
Using a wide range of nebula pressures and core masses we can calculate
the possibly existing envelope-core combinations in hydrostatic
equilibrium. Assuming that all such states are equally likely, the
relative frequency of planetary masses (core+envelope mass)
corresponds to a distribution function of planet masses.

\subsection{Calculation of the envelope structures}
\label{sec:calc-envel-struct}
Before we discuss the prediction method in detail, we'll describe how
the individual planet-envelope structures are calculated.

Each "planet candidate" consists of a solid core of fixed
density\footnote{we use the value of $\rho=5500\,\mathrm{kg}
\,  \mathrm{m}^{-3}$} that is embedded in a nebula of a particular
pressure. The envelope structure is determined by the well-known
equations of stellar structure \citep[e.g.][]{kippenhahn}. 
These, we calculate in radial symmetry and neglect rotation. The
effect of rotation is
negligible in all but the most extremely rotating cases
\citep{1993DissGoetz}. 

A constant infall of planetesimals onto the core releases gravitational
energy that is transported through the envelope by either radiation or
convection. We use the diffusion approximation for the radiative
energy transport and zero entropy convection.

The properties of the envelope are determined by the equation of state
by \citet{1995ApJS...99..713Sv}. Rosseland-mean opacities
$\kappa(\rho,T)$ are interpolated from  
a combined table: Opacities include Rosseland-mean dust opacities
from \citet[][$\lg T \leq 2.3$]{1985Icar...64..471P},
\citet{1994ApJ...437..879A} values 
in the molecular range, and \citet{1990ADNDT..45..209W} Los Alamos
high temperature opacities.

The planet extends out to the hill-radius where the pressure of
envelope and nebula are set to be equal.

As discussed in \citet{broegHD149026b}, the problem is fully specified
when the following six quantities are specified:\\
The
\begin{enumerate}
\item core mass $M_\mathrm{core}$,\label{item:1}
\item pressure at the core $P_\mathrm{core}$,\label{item:2}
\item mass of the host star  $M_\ast$,\label{item:3}
\item semi-major-axis of the planet $a$, \label{item:4}
\item nebula temperature $T_\mathrm{neb}$, and the \label{item:5}
\item planetesimal accretion rate $\dot M$.\label{item:6}
\end{enumerate}
The parameters \ref{item:1} and \ref{item:2} are our independent
parameters. By varying these two independent parameters, we can
determine all possible hydrostatic envelope solutions for a given
"location". A "location" is determined by the parameters
\ref{item:3}-\ref{item:6}. They give the environmental conditions of
the proto-planet.

Parameters 
\ref{item:3},\ref{item:4}, and \ref{item:5} are determined by the host
star and the location of the planet. The nebula temperature can be calculated
in thermal equilibrium with the star: 
\begin{equation}
  \label{eq:nebT}
  T_\mathrm{neb} = 280\cdot \left(\frac{a}{1 \mathrm{AU}}\right)^{-1/2} \left(\frac{L_\ast}{L_{\sun}}\right)^{1/4} \, \mathrm{K}
\end{equation}
with $L_\ast$ the luminosity of the planet host star and $L_{\sun}$ the solar
luminosity
\citep[see][]{1981PThPS..70...35H,1985prpl.conf.1100H}. This implies a
passive disk, i.e. no viscous heating and assumes
that the nebula is optically thin.

The only remaining free parameter is the planetesimal accretion rate
$\dot M$. Proper values in agreement
with planetesimal theory range from
 $\dot M=10^{-2}\,\mathrm{M}_{\earth}\, \mathrm{a}^{-1}$(note that a
   stands for one year here, not the semi-major-axis) very close to the star to $\dot
 M=10^{-6}\,\mathrm{M}_{\earth}\, \mathrm{a}^{-1}$ at Jupiter distances\footnote{The
   high value corresponds to a orbital distance of
   $a=0.04$\,AU,  
 particle-in-a-box planetesimal accretion theory with a gravitational
 enhancement factor $F_\mathrm{g}\approx 20$,  a 
 minimum mass solar nebula
 \citep{1981PThPS..70...35H,1985prpl.conf.1100H} and Jupiter mass objects.}. 

For a detailed description of the equations and boundary conditions, see
\citet{broegHD149026b}.

\subsection{Calculating a mass spectrum for a fixed location}
\label{sec:calc-mass-spectr}

\subsubsection{A set of solutions for a fixed location: the manifold}
\label{sec:set-solutions-fixed}
\begin{figure}
  \centering
\includegraphics[width=\columnwidth]{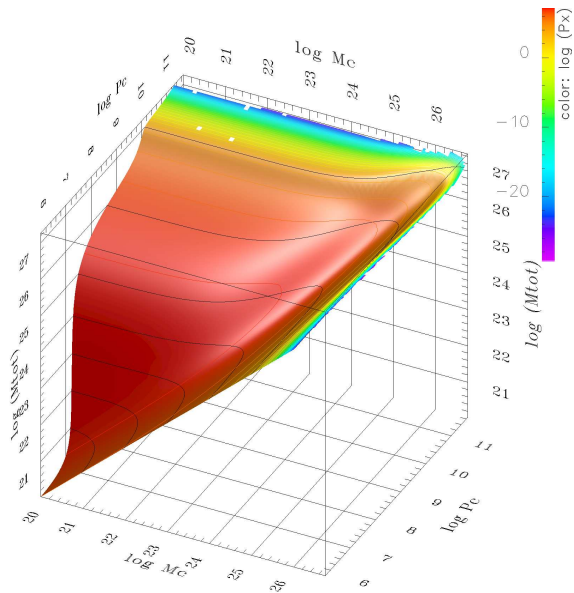}
  \caption{Manifold of proto-planet masses for a four day orbit around
    a solar type host star. The accretion parameter is set to $\dot
 M=10^{-4}\,\mathrm{M}_{\earth}\, \mathrm{a}^{-1}$. The total proto-planet mass as a
 function of the parameters \mc, \pc. All axes are logarithmic. The
 results are connected to show a 3-dimensional surface. This surface
 is colored using the color-map shown in the upper right corner
 mapping the outside pressure $P_\mathrm{x}$ in a logarithmic
 way. This shows that the outside pressure is varied from $\approx
 10^4 to \approx 10^{-30}\,\mathrm{Pa}$. Everything with higher nebula
 pressures is by construction gravitationally unstable, lower
 pressures correspond to vacuum and cannot be calculated here properly
 because we do not calucate an atmosphere. This plot covers more than 6 magnitudes in  \mc and \mtot, and 5
 magnitudes in \pc.} 
  \label{fig:1}
\end{figure}

For a given location as defined by the parameters
\ref{item:3}-\ref{item:6} in section~\ref{sec:calc-envel-struct} we
can calculate all hydrostatic equilibrium solutions to the equations
of stellar structure. One such set of solutions covering a wide range
in the \mc,\pc-plane we term, following \citet{2005A&A...440.1183P}, a "manifold" or "solution manifold" for
the given location. Each manifold contains, once calculated, \emph{all
envelope structures} that can possibly exist hydro-statically inside
any nebula at the given location. One example for such a manifold is
given in figure~\ref{fig:1}. It shows the total mass of the
proto-planet as a function of the parameters \mc and \pc.

\subsubsection{Deriving the mass spectrum from a manifold}
\label{sec:deriv-mass-spectr}

Having calculated a manifold, we now make the following assumptions:
\begin{enumerate}
\item All equilibria are equally probable.\label{item:7}
\item All equilibria are stable and can be dynamically reached.\label{item:8}
\end{enumerate}

Now  we can --
quite in analogy to statistical mechanics -- derive a distribution
function for various properties of the proto-planets at that
"location". The quantity we are interested in is the mass of the
planet. By quite literally counting off the occuring masses in the
manifold we can derive what we call the "mass spectrum": The relative
frequency of planet masses at this location\footnote{In order to
  produce a histogram of continuous data, the data have to be binned
  to a fixed bin size. We chose a logarithmic binning with a bin size
  of 0.05 dex.}.

For this application of the manifold, it is important not to choose a
certain range of core masses and 
core pressures implicitly by choice of a scale. Therefore, we chose a
scale-free distribution 
to sample the parameter space
in $M_\mathrm{core}$ and $P_\mathrm{core}$. In this way no core mass
or core pressure is selected and all values are treated alike. The only
scale-free distribution is a power-law distribution, or
correspondingly a log-equidistant sampling of the parameter space.
 Obviously,
if there were a dominant scale, e.g. in the core mass, this would
drastically change the outcome as compared to our scale-free set-up.

\begin{figure}
  \centering
  \includegraphics[width=\columnwidth]{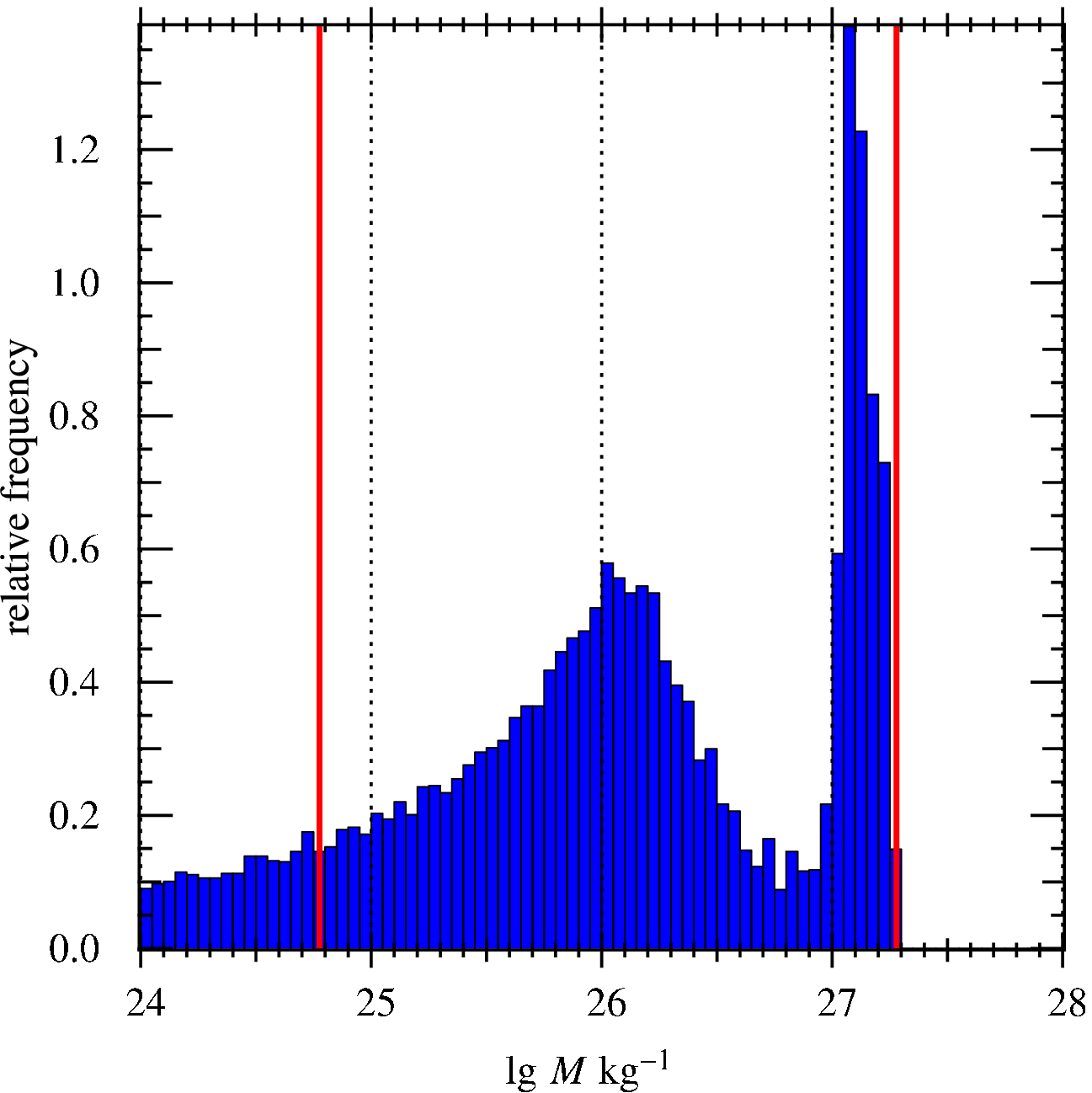}
  \caption{Mass spectrum of proto-planet masses for a four day orbit around
    a solar type host star. The accretion parameter is set to $\dot
 M=10^{-4}\,\mathrm{M}_{\earth}\, \mathrm{a}^{-1}$. The area is normalized to 1. The red
lines mark the value of  1\,$\mathrm{M}_{\earth}$ and 1\,\mjup. The mass-distribution for 1\,\msun at 4\,d is clearly dominated by two
peaks, one at 
$\approx17\, \mearth$ and another at $\approx210\,\mearth$ or
$\approx0.6\,\mjup$. }
  \label{fig:2}
\end{figure}

The result for the same location as figure~\ref{fig:1} is shown in
figure~\ref{fig:2}. This mass spectrum is derived based on assumptions
(i,ii). Therefore, either dynamical
instability of some parts of the manifold, or quasistatic contraction
with significant mass gain will change the observed mass spectrum. On
the other hand, agreement of the mass spectra with observation would
be a strong indication for the validity of the assumptions (i,ii).

\section{planet prediction results}
\label{sec:plan-pred-results}

\subsection{Manifold survey}
\label{sec:manifold-survey}

We have calculated manifolds and corresponding mass spectra for a wide
range of locations by varying the following three parameters:
\begin{description}
\item[${(4)\,T_\textrm{orb}:}$]$1,\; 4,\; 16,\; 64$ days
\item[${(4)\,M_{\ast}:}$]$2,\; 1,\; 0.8,\; 0.4\, \msun$;
  $L_{\ast}=16,\; 1,\; 0.42,\; 0.04\, \lsun$.\footnote{This corresponds roughly to
  spectral types A2, G2, K1, or. M2. Luminosities of the host star
  are assigned to the 
  masses following \citet{Gray92}.}
\item[${(3)\,\mdot:}$]  $10^{-2},\; 10^{-4},\;
  10^{-6}\,\mearth\,\textrm{a}^{-1}$.
\end{description}
This results in a 3-dimensional grid of locations, a total of 48
manifolds. This is the first complete survey of hydrostatic
proto-planets in close orbits. The full set of 
results can be seen in \citet{DissBroegChristopher} and is also available on-line
at \citet{corotWeb}.
This survey -- named Corot survey Mark 1 v1.1 -- revealed a large diversity of mass spectra in the range
from 1 to 64 days orbital period. The host star mass also has  large
impact on the mass distributions.

\subsection{Statistical properties -- Three classes of gas giants}
\label{sec:statistical-prop}
As stated in section~\ref{sec:deriv-mass-spectr} a manifold can be
used to determine the mass spectrum using the following two
hypotheses: 1) all equilibria are equally probable and 2) they can be
dynamically reached, i.e. there exists a track from some set of initial
conditions to each state. Using these hypotheses we can derive several
interesting properties of the giant planets.

One major result of this survey is the fact that all mass spectra for
close-in orbits
exhibit two peaks. This is so for all tested values of \mdot.  Moving to larger orbital 
distances, these peaks move closer together and eventually merge into one
peak. For a solar type host star, this happens at an orbital period of
around 
16 days.

The full set of mass spectra of our survey
leads to the grouping of the planets into three classes:
\begin{description}
\item[\textbf{Class G}]  \emph{Extremely hot gas giants}\footnote{German
    \emph{\textbf{G}anz heiß}} reside very close to the host star.
  Their surface 
  temperature is above dust sublimation temperature. Planets in this
  class have a very 
  large upper mass limit\footnote{derived as the largest occuring masses in the mass
  spectrum}. For solar type host stars,
  the mass limit is roughly at 2.5\,\mjup.\footnote{This value is
    strongly dependent on \mdot. The given value corresponds to
    $\mdot=10^{-4}\,\mearth\,\textrm{a}^{-1}$. It is 5.3 and 0.8\,\mjup
  for an $\mdot$ of $10^{-2}$ and $10^{-6}$ respectively.} For host stars of
  2\,\msun, this limit is extended up to 6\,\mjup. More massive
  planets should not exist this close to the star. In addition, we
  expect a very large quantity of so-called hot Neptunes with masses
  around 16\,\mearth corresponding to a large second peak in the mass
  spectrum. 
\item[\textbf{Class H}]  \emph{\textbf{H}ot gas giants} reside in-between
  the classes G and J. Their surface temperature is below dust
  sublimation and they are close enough to their host star so that the
  mass spectra still show two distinct peaks. We expect them to be less massive than
  1\,\mjup (for a 1\,\msun host star).
\item[\textbf{Class J}] \emph{\textbf{J}upiter-like gas giants} show only
  one peak in their mass spectrum.  Class~J planets can  be much more massive
  than the classes G and H because the equilibria can gain significant
  amounts of mass by quasi-static contraction while the nebula is
  still present
   (see section \ref{sec:discussion}).
\end{description}
For a solar type star, the boundaries between the groups G,H and H,J
are at 4 days 
and 16 days  orbital period, respectively.\footnote{The 16 day boundary depends
strongly on the host star mass and the planetesimal accretion rate. 16 days
correspond to a 1\,\msun star and a low accretion rate ($\dot
 M=10^{-6}\,\mathrm{M}_{\earth}\, \mathrm{a}^{-1}$). For a slightly higher accretion
 rate ($\dot
 M=10^{-4}\,\mathrm{M}_{\earth}\, \mathrm{a}^{-1}$), the two peaks in the mass spectrum
 merge around  32 days orbital period.}

\subsection{Discussion}
\label{sec:discussion}
As discussed in the last section, the transition from class H to J is
marked by the merger of the two peaks in the mass spectrum. We have
performed both  isothermal linear instability analysis 
\citep{2005Schoenke} and isothermal non-linear
instability analysis 
\citep{2005Pecnik} of a number of manifolds. These calculations
suggest a fundamental change in 
dynamical properties that coincides (or is caused by) the merger of
the peaks: In the
one-peak case,
entire regions in the manifold appear to be unstable owing to
transitions between two states of similar mass.

Another change in  behaviour at the merge position can be derived from the manifolds directly: At
large orbital distances, there is a well-defined critical core mass
beyond which no more static solutions exist inside a nebula. The value
of the critical core mass depends only weakly on nebula pressure. This
leads to the explanation why Jupiter and Saturn appear to have similar cores.

 At
small orbital distances, however, the critical core-mass becomes very
strongly dependant on nebula pressure. In consequence, it is always
possible to find a nebula pressure where the core is sub-critical,
i.e. where a static solution exists. This renders the concept of a
critical core mass meaningless for classes H and G.
Following the above line of reasoning for close-in planets, it follows
that no significant mass gain is expected by the disappearance of the nebula.
Therefore the calculated mass spectrum of classes G and H could be
very much like the observed mass spectrum in this regime.

These considerations have  been tested in a small number of 
calculations using full radiation hydrodynamic planet formation.
One such case is the planet
 HD\,149026\,b at a distance of 0.042\,AU corresponding to an orbital
 period of 2.87 days. Our calculations show a completely hydrostatic
 evolution \citep[see][]{broegHD149026b} without significant mass gain
 in the final phase. The final planet has the same mass as the
 equilibrium configuration in the manifold.

\subsection{First comparison with observations}
\label{sec:first-comp-with}
\citet{2005Sci...310..251S} detect a paucity of high-mass planetary
companions with orbital periods shorter than $\sim40$\,days. This is
in agreement with our separation in upper mass limit classes (G\&H) and
the J class without a strict upper mass limit.

\citet{2005ApJ...623..472G} went  a step further dividing the
observed exo-planets into very hot (VHJ) and hot Jupiters (HJ) with a
dividing line at 3 days orbital period. They observe that the VHJ
exhibit higher masses than the HJ; Specifically, the VHJ masses are larger than
1\,\mjup. This is in perfect agreement with our separation into groups
G \& H and the upper mass limit of 1\,\mjup for group H.

\begin{figure}
   \centering
   \includegraphics[width=\columnwidth]{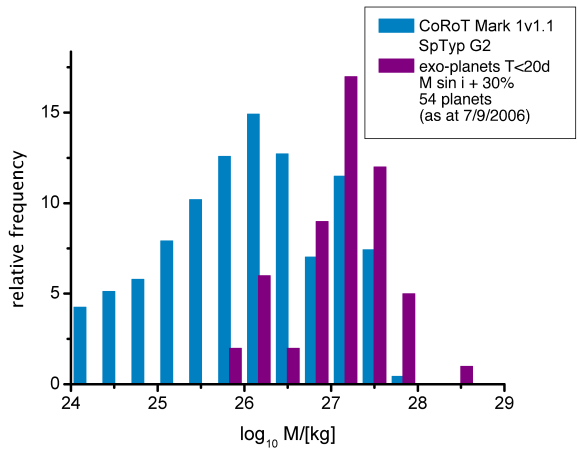}
   \caption{Predicted and observed mass distributions \citep[Observed
     exo-planets from][on
     July 7, 2006]{Schneider}. For this prediction we used only solar
     type host stars. The high-mass peak is in good agreement
   with the observed data. Only the very high data point is not
   reproduced but it might be a brown dwarf with a different formation
   mechanism. Including higher mass host stars would produce slightly
   higher upper masses in our prediction.  The lower-mass peak of our
   calculations can already be noticed in the observed data but these
   planets are at the detection limits. We expect this part of the observed
   distribution to grow as the instruments become more sensitive. The
   high-mass end, on the other hand, should be complete for the
   observed stars.}
   \label{fig:3}
 \end{figure}

As a final comparison of our method to observations we performed a
direct "prediction" for the  host stars of today's
exo-planets. Because of the class properties, nameley the stability of
equilibrium states for classes G \& H and the lack thereof for the J
class, we only compare our predicted masses at
orbital periods less than 20 days. At the time of our analysis, 54
exo-planets fell in that regime. For the prediction we assumed all
host stars to be solar.\footnote{Most exo-planet host stars to date
  are of solar metallicity.} 
The reference mass spectrum was obtained by binning the masses of
these 54 exo-planets in 0.3 dex mass bins. As
 most of the detected exo-planets have been observed using the radial
 velocity technique, we added a 30\% correction to the observed planet
 masses  as is statistically
 expected.  This we compared to our theoretical mass spectrum:
For each detected exo-planet we computed a mass spectrum using the
corresponding orbital distance. All such mass spectra were added and
binned according to the reference mass spectrum.\footnote{More
  precisely, we used the 
  following procedure to obtain the theoretical mass spectrum: 1) The
  detected exo-planets are grouped into bins of orbital period
  (0--2,2--8,8--20 days). 2) The computed mass spectra for the 
  planetesimal accretion rates $\dot 
 M=10^{-2}$ and $10^{-4}\,\mathrm{M}_{\earth}\, \mathrm{a}^{-1}$ at a given
  orbital period (1, 4, 16 days) are multiplied with the number of
  exo-planets in the corresponding period range and added together. 3)
The resulting mass spectrum is then normalized and binned like 
the reference spectrum.}  The resulting mass spectrum is compared to
the reference spectrum in figure~\ref{fig:3}. Please note that our
calculated mass spectrum can only be expected to reproduce the
observed one if there are no phases of quasi-static contraction while
the nebula still exists, i.e. while there is a significant mass
reservoir for the planet (this is
the case for HD 149026b).

\section{Conclusion}
\label{sec:conclusion}
We have presented a new method to predict the mass distribution of gas
giant planets that analyses all possible hydrostatic
equilibria. It has the advantage of not needing the
nebula density as input. Only nebula temperature and planetesimal
accretion rate must be known. In
a passive disk, this can be easily
approximated using the host star properties. This leaves the
planetesimal accretion rate as the only free parameter.

Using our new method, we are able to split the giant
planets into three 
classes G, H, and J which have distinct properties (see
section~\ref{sec:statistical-prop}). We compare these properties
to the observed mass distribution of the exo-planets and find good
agreement. We also produce a mass distribution for the exo-planet host
stars having close-in planets and can reproduce the observed mass distribution.
The agreement with observations is a strong argument that the
equilibria are indeed dominating the formation process of close-in
planets and that a large variety of proto-planetary nebulae is in existence.

To produce a prediction for \corot\ star fields, the existing mass spectra
have to be averaged according to the distribution of stars in the
\corot\ fields and a concept has to be developed to determine the
relative planet abundance of planets at different orbital
distances. So far we can only predict planetary mass distributions at
given orbital distances.

\section*{Acknowledgements}
This research was supported in part by DLR
project number 
50-OW-0501.

\bibliographystyle{mn}
\bibliography{lit/Literatur,lit/submitted}

\def\aj{AJ} \def\araa{ARA\&A} \def\apj{ApJ} \def\apjl{ApJ} \def\apjs{ApJS}
  \def\ao{Appl.~Opt.} \def\apss{Ap\&SS} \def\aap{A\&A} \def\aapr{A\&A~Rev.}
  \def\aaps{A\&AS} \def\azh{AZh} \def\baas{BAAS} \def\jrasc{JRASC}
  \def\memras{MmRAS} \def\mnras{MNRAS} \def\pra{Phys.~Rev.~A}
  \def\prb{Phys.~Rev.~B} \def\prc{Phys.~Rev.~C} \def\prd{Phys.~Rev.~D}
  \def\pre{Phys.~Rev.~E} \def\prl{Phys.~Rev.~Lett.} \def\pasp{PASP}
  \def\pasj{PASJ} \def\qjras{QJRAS} \def\skytel{S\&T} \def\solphys{Sol.~Phys.}
  \def\sovast{Soviet~Ast.} \def\ssr{Space~Sci.~Rev.} \def\zap{ZAp}
  \def\nat{Nature} \def\iaucirc{IAU~Circ.} \def\aplett{Astrophys.~Lett.}
  \def\apspr{Astrophys.~Space~Phys.~Res.}
  \def\bain{Bull.~Astron.~Inst.~Netherlands} \def\fcp{Fund.~Cosmic~Phys.}
  \def\gca{Geochim.~Cosmochim.~Acta} \def\grl{Geophys.~Res.~Lett.}
  \def\jcp{J.~Chem.~Phys.} \def\jgr{J.~Geophys.~Res.}
  \def\jqsrt{J.~Quant.~Spec.~Radiat.~Transf.}
  \def\memsai{Mem.~Soc.~Astron.~Italiana} \def\nphysa{Nucl.~Phys.~A}
  \def\physrep{Phys.~Rep.} \def\physscr{Phys.~Scr}
  \def\planss{Planet.~Space~Sci.} \def\procspie{Proc.~SPIE} \let\astap=\aap
  \let\apjlett=\apjl \let\apjsupp=\apjs \let\applopt=\ao \def\aj{AJ}
  \def\araa{ARA\&A} \def\apj{ApJ} \def\apjl{ApJ} \def\apjs{ApJS}
  \def\ao{Appl.~Opt.} \def\apss{Ap\&SS} \def\aap{A\&A} \def\aapr{A\&A~Rev.}
  \def\aaps{A\&AS} \def\azh{AZh} \def\baas{BAAS} \def\jrasc{JRASC}
  \def\memras{MmRAS} \def\mnras{MNRAS} \def\pra{Phys.~Rev.~A}
  \def\prb{Phys.~Rev.~B} \def\prc{Phys.~Rev.~C} \def\prd{Phys.~Rev.~D}
  \def\pre{Phys.~Rev.~E} \def\prl{Phys.~Rev.~Lett.} \def\pasp{PASP}
  \def\pasj{PASJ} \def\qjras{QJRAS} \def\skytel{S\&T} \def\solphys{Sol.~Phys.}
  \def\sovast{Soviet~Ast.} \def\ssr{Space~Sci.~Rev.} \def\zap{ZAp}
  \def\nat{Nature} \def\iaucirc{IAU~Circ.} \def\aplett{Astrophys.~Lett.}
  \def\apspr{Astrophys.~Space~Phys.~Res.}
  \def\bain{Bull.~Astron.~Inst.~Netherlands} \def\fcp{Fund.~Cosmic~Phys.}
  \def\gca{Geochim.~Cosmochim.~Acta} \def\grl{Geophys.~Res.~Lett.}
  \def\jcp{J.~Chem.~Phys.} \def\jgr{J.~Geophys.~Res.}
  \def\jqsrt{J.~Quant.~Spec.~Radiat.~Transf.}
  \def\memsai{Mem.~Soc.~Astron.~Italiana} \def\nphysa{Nucl.~Phys.~A}
  \def\physrep{Phys.~Rep.} \def\physscr{Phys.~Scr}
  \def\planss{Planet.~Space~Sci.} \def\procspie{Proc.~SPIE} \let\astap=\aap
  \let\apjlett=\apjl \let\apjsupp=\apjs \let\applopt=\ao
\begin{thebibliography}{42}
\expandafter\ifx\csname natexlab\endcsname\relax\def\natexlab#1{#1}\fi

\bibitem[{{Alexander} \& {Ferguson}(1994)}]{1994ApJ...437..879A}
{Alexander} D.~R., {Ferguson} J.~W., 1994, \apj, 437, 879

\bibitem[{{Alibert} {et~al.}(2005{\natexlab{a}}){Alibert}, {Mordasini}, {Benz},
  \& {Winisdoerffer}}]{2005A&A...434..343A}
{Alibert} Y., {Mordasini} C., {Benz} W., {Winisdoerffer} C.,
  2005{\natexlab{a}}, \aap, 434, 343

\bibitem[{{Alibert} {et~al.}(2005{\natexlab{b}}){Alibert}, {Mousis}, \&
  {Benz}}]{2005ApJ...622L.145A}
{Alibert} Y., {Mousis} O., {Benz} W., 2005{\natexlab{b}}, \apjl, 622, L145

\bibitem[{{Alibert} {et~al.}(2005{\natexlab{c}}){Alibert}, {Mousis},
  {Mordasini}, \& {Benz}}]{2005ApJ...626L..57A}
{Alibert} Y., {Mousis} O., {Mordasini} C., {Benz} W., 2005{\natexlab{c}},
  \apjl, 626, L57

\bibitem[{{Baglin} \& {The COROT Team}(1998)}]{1998IAUS..185..301B}
{Baglin} A., {The COROT Team}, 1998, in IAU Symp. 185: New Eyes to See Inside
  the Sun and Stars, {Deubner} F.-L., {Christensen-Dalsgaard} J., {Kurtz} D.,
  eds., pp. 301--+

\bibitem[{{Benz} {et~al.}(2006){Benz}, {Mordasini}, {Alibert}, \&
  {Naef}}]{2006tafp.conf...24B}
{Benz} W., {Mordasini} C., {Alibert} Y., {Naef} D., 2006, in Tenth Anniversary
  of 51 Peg-b: Status of and prospects for hot Jupiter studies, {Arnold} L.,
  {Bouchy} F., {Moutou} C., eds., pp. 24--34

\bibitem[{{Bodenheimer} \& {Pollack}(1986)}]{1986Icar...67..391B}
{Bodenheimer} P., {Pollack} J.~B., 1986, Icarus, 67, 391

\bibitem[{{Bord{\'e}} {et~al.}(2003){Bord{\'e}}, {Rouan}, \&
  {L{\'e}ger}}]{2003A&A...405.1137B}
{Bord{\'e}} P., {Rouan} D., {L{\'e}ger} A., 2003, \aap, 405, 1137

\bibitem[{{Borucki} \& {Summers}(1984)}]{1984Icar...58..121B}
{Borucki} W.~J., {Summers} A.~L., 1984, Icarus, 58, 121

\bibitem[{{Boss}(2002)}]{2002ApJ...576..462B}
{Boss} A.~P., 2002, \apj, 576, 462

\bibitem[{{Broeg}(2006{\natexlab{a}})}]{corotWeb}
{Broeg} C., 2006{\natexlab{a}}, Corot survey,
  http://www.astro.uni-jena.de/corot/

\bibitem[{{Broeg}(2006{\natexlab{b}})}]{DissBroegChristopher}
---, 2006{\natexlab{b}}, PhD thesis, Friedrich-Schiller-Universit{\"a}t Jena,
  Germany

\bibitem[{Broeg \& Wuchterl(2006)}]{broegHD149026b}
Broeg C., Wuchterl G., 2006, The formation of hd 149026b, {MNRAS}, accepted
  2007 Jan 9, astro-ph/0701340

\bibitem[{{Charbonneau} {et~al.}(2000){Charbonneau}, {Brown}, {Latham}, \&
  {Mayor}}]{2000ApJ...529L..45C}
{Charbonneau} D., {Brown} T.~M., {Latham} D.~W., {Mayor} M., 2000, \apjl, 529,
  L45

\bibitem[{COROT(2006)}]{corot}
COROT, 2006, Corot, http://smsc.cnes.fr/COROT/

\bibitem[{{Gaudi} {et~al.}(2005){Gaudi}, {Seager}, \&
  {Mallen-Ornelas}}]{2005ApJ...623..472G}
{Gaudi} B.~S., {Seager} S., {Mallen-Ornelas} G., 2005, \apj, 623, 472

\bibitem[{{Goldreich} {et~al.}(2004){Goldreich}, {Lithwick}, \&
  {Sari}}]{2004ARA&A..42..549G}
{Goldreich} P., {Lithwick} Y., {Sari} R., 2004, \araa, 42, 549

\bibitem[{{G{\"o}tz}(1989)}]{1993DissGoetz}
{G{\"o}tz} M., 1989, PhD thesis, Univ. Heidelberg

\bibitem[{Gray(1992)}]{Gray92}
Gray D., 1992, The observation and analysis of stellar atmospheres, 2nd edn.
  Cambridge University Press, Cambridge

\bibitem[{{Hayashi}(1981)}]{1981PThPS..70...35H}
{Hayashi} C., 1981, Progress of Theoretical Physics Supplement, 70, 35

\bibitem[{{Hayashi} {et~al.}(1977){Hayashi}, {Nakazawa}, \&
  {Adachi}}]{1977PASJ...29..163H}
{Hayashi} C., {Nakazawa} K., {Adachi} I., 1977, \pasj, 29, 163

\bibitem[{{Hayashi} {et~al.}(1979){Hayashi}, {Nakazawa}, \&
  {Mizuno}}]{1979E&PSL..43...22H}
{Hayashi} C., {Nakazawa} K., {Mizuno} H., 1979, Earth and Planetary Science
  Letters, 43, 22

\bibitem[{{Hayashi} {et~al.}(1985){Hayashi}, {Nakazawa}, \&
  {Nakagawa}}]{1985prpl.conf.1100H}
{Hayashi} C., {Nakazawa} K., {Nakagawa} Y., 1985, in Protostars and Planets II,
  pp. 1100--1153

\bibitem[{Kippenhahn \& Weigert(1990)}]{kippenhahn}
Kippenhahn R., Weigert A., 1990, Stellar Structure and Evolution, Harwit M.,
  Kippenhahn R., Trimble V., Zahn J.-P., eds., Astronomy and Astrophysics
  Library. Springer-Verlag, Berlin, Heidelberg

\bibitem[{{Lissauer}(1993)}]{1993ARA&A..31..129L}
{Lissauer} J.~J., 1993, \araa, 31, 129

\bibitem[{{Mayor} \& {Queloz}(1995)}]{1995Natur.378..355M}
{Mayor} M., {Queloz} D., 1995, \nat, 378, 355

\bibitem[{{Mizuno}(1980)}]{mizuno1980}
{Mizuno} H., 1980, Progress of Theoretical Physics, 64, 544

\bibitem[{{Pe{\v c}nik} \& {Wuchterl}(2005)}]{2005A&A...440.1183P}
{Pe{\v c}nik} B., {Wuchterl} G., 2005, \aap, 440, 1183

\bibitem[{{{Pe{\v c}nik}, B.}(2005)}]{2005Pecnik}
{{Pe{\v c}nik}, B.}, 2005, Ph.d.thesis, Ludwig-Maximillians-Universit{\"a}t
  M{\"u}nchen, M{\"u}nchen, Germany

\bibitem[{{Pollack} {et~al.}(1996){Pollack}, {Hubickyj}, {Bodenheimer},
  {Lissauer}, {Podolak}, \& {Greenzweig}}]{1996Icar..124...62P}
{Pollack} J.~B., {Hubickyj} O., {Bodenheimer} P., {Lissauer} J.~J., {Podolak}
  M., {Greenzweig} Y., 1996, Icarus, 124, 62

\bibitem[{{Pollack} {et~al.}(1985){Pollack}, {McKay}, \&
  {Christofferson}}]{1985Icar...64..471P}
{Pollack} J.~B., {McKay} C.~P., {Christofferson} B.~M., 1985, Icarus, 64, 471

\bibitem[{{Rouan} {et~al.}(2000){Rouan}, {Baglin}, {Barge}, {Bord{\'e}},
  {Deleuil}, {L{\'e}ger}, {Schneider}, \& {Vuillemin}}]{2000dais.conf..221R}
{Rouan} D., {Baglin} A., {Barge} P., {Bord{\'e}} P., {Deleuil} M., {L{\'e}ger}
  A., {Schneider} J., {Vuillemin} A., 2000, in ESA SP-451: Darwin and Astronomy
  : the Infrared Space Interferometer, {Sch{\"u}rmann} B., ed., pp. 221--+

\bibitem[{{Santos} {et~al.}(2005){Santos}, {Benz}, \&
  {Mayor}}]{2005Sci...310..251S}
{Santos} N.~C., {Benz} W., {Mayor} M., 2005, Science, 310, 251

\bibitem[{{Saumon} {et~al.}(1995){Saumon}, {Chabrier}, \& {van
  Horn}}]{1995ApJS...99..713Sv}
{Saumon} D., {Chabrier} G., {van Horn} H.~M., 1995, \apjs, 99, 713

\bibitem[{{Schneider}(2006)}]{Schneider}
{Schneider} J., 2006, The extrasolar planets encyclopedia, http://exoplanet.eu/

\bibitem[{{{Sch{\"o}nke}, J.}({2005})}]{2005Schoenke}
{{Sch{\"o}nke}, J.}, {2005}, {Diplomarbeit}, {Friedrich Schiller
  Universit{\"a}t Jena}

\bibitem[{{Weiss} {et~al.}(1990){Weiss}, {Keady}, \&
  {Magee}}]{1990ADNDT..45..209W}
{Weiss} A., {Keady} J.~J., {Magee} N.~H., 1990, Atomic Data and Nuclear Data
  Tables, 45, 209

\bibitem[{{Wetherill} \& {Stewart}(1989)}]{1989Icar...77..330W}
{Wetherill} G.~W., {Stewart} G.~R., 1989, Icarus, 77, 330

\bibitem[{{Wuchterl}(1990)}]{1990A&A...238...83W}
{Wuchterl} G., 1990, \aap, 238, 83

\bibitem[{{Wuchterl}(1991{\natexlab{a}})}]{1991Icar...91...39W}
---, 1991{\natexlab{a}}, Icarus, 91, 39

\bibitem[{{Wuchterl}(1991{\natexlab{b}})}]{1991Icar...91...53W}
---, 1991{\natexlab{b}}, Icarus, 91, 53

\bibitem[{{Wuchterl} {et~al.}(2000){Wuchterl}, {Guillot}, \&
  {Lissauer}}]{2000prpl.conf.1081W}
{Wuchterl} G., {Guillot} T., {Lissauer} J.~J., 2000, Protostars and Planets IV,
  1081

\end{thebibliography}

\label{lastpage}
\end{document}